\documentclass[epjCONF,columns]{svjour}
\usepackage{graphics}
\usepackage[varg]{txfonts} 
\usepackage[latin1]{inputenc}
\def\met{\ensuremath{E_{\mathrm{T}}^{\mathrm{miss}}}} 

\session-title{2011 Hadron Collider Physics Symposium}
\begin{document}
%
\title{Measurement of the $b$-jet cross-section with associated vector boson production with the ATLAS experiment at the LHC}
\author{Heather M. Gray\thanks{\email{heather.gray@cern.ch}} on behalf of the ATLAS Collaboration}
\institute{CERN}
\abstract{
A measurement of the cross-section for vector boson production in association with jets containing b-hadrons is presented using 35 pb$^{-1}$ of data from the LHC collected by the ATLAS experiment in 2010. Such processes are not only important tests of pQCD but also large, irreducible backgrounds to searches such as a low mass Higgs boson decaying to pairs of $b$-quarks when the Higgs is produced in association with a vector boson. Theoretical predictions of the V+$b$ production rate have large uncertainties and previous measurements have reported discrepancies. Cross-sections measured using in the electron and muon channels will be shown. Comparisons will be made to recent theoretical predictions at the next-to-leading order in $\alpha_S$.
} 
\maketitle
%
Vector boson production in association with jets (V+jets) is often used as a testing ground for perturbative QCD calculations. Despite substantial progress in the understanding and modelling of inclusive jet production in vector boson events, less study has been made of heavy-flavour jet production. V+jets is also an important background to many searches at the LHC. These include searches for the Higgs boson produced in association with a vector boson and with the Higgs decaying to a pair of $b$-quarks~\cite{Hbb} or supersymmetric models with $b$-quarks in the final state\cite{SBottom}. Significant progress has been made recently in increasing the accuracy of theoretical calculations~\cite{CampWb,FFSWb}. Challenges to theoretical calculations stem from the non-necessarily negligible $b$-quark mass and the interplay between contributions from $b$-quark production in initial and final states. We discuss results of a measurement of the W and Z + $b$-jet cross-sections made by the ATLAS experiment using 35~pb$^{-1}$ of LHC data collected in 2010~\cite{refZb,refWb}. Both measurements are challenging, but in different respects: the Z+$b$-jet is statistically limited by its small cross-section, while the W+$b$-jet process has a larger cross-section, but also larger backgrounds resulting in a low signal to background ratio.

The V+$b$-jet cross-sections are measured in a fiducial phase space, by correcting the results back to the particle level while accounting for all detector effects. The cuts applied to define the fiducial cross-section, closely matching the experimental acceptance, are listed in Table~\ref{tab:1}. The lepton momentum includes the energy from photons radiated within a cone of radius 0.1. The neutrino $p_T$ is used as the transverse missing energy. Particle level jets are reconstructed using the anti-$k_T$ algorithm using all stable particles ($\tau > 10$~ps).
A $b$-jet is defined as a jet with a $b$-hadron with $p_T > 5$~GeV with an angular separation from the jet $\Delta R < 0.3$. 

The Z+$b$-jet cross-section is measured inclusively for events containing at least one $b$-jet. Due to the larger W+$b$-jet cross-section a differential measurement is made for events containing either one or two $b$-jets, but events containing additional jets are vetoed to control backgrounds. The Z+$b$-jet cross-section is measured at the jet level, such that events with two jets enter the distributions twice. The W+$b$-jet cross-section, is measured at the event level, with each event entering each distribution once.

The ATLAS detector consists of an inner tracking system ($|\eta| < 2.5$) surrounded by a 2~T super-conducting sole\-noid, electromagnetic and hadronic calorimeters ($|\eta| < 4.9$), and a muon spectrometer ($|\eta| < 2.7$) ~\cite{DetPaper}. Events are collected using single electron or muon high $p_T$ triggers and required to contain at least one reconstructed primary vertex. Leptons are required to have $p_T > 20$~GeV with electrons within $|\eta| < 2.47$ and muons within $|\eta| < 2.4$. Events containing a Z boson are selected by requiring two isolated leptons of the same flavour but with opposite charge and with the invariant mass of the leptons to be consistent with the Z boson mass: $76 < m_{ll} < 106$~GeV. Events containing a W boson are selected by requiring exactly one isolated lepton with missing transverse energy ($\met > 25$~GeV) and transverse mass, $m_T^W = \sqrt{2 p_T^\ell p_T^\nu ( 1 - \mathrm{cos}(\phi^{ell} - \phi^\nu))}$, consistent with that of a W boson ($m_T^W > 40$~GeV).

Jets are reconstructed using the anti-$k_T$ algorithm with a radius parameter $R = 0.4$ and required to have $p_T > 25$~GeV. Jets with an angular separation $\Delta R < 0.5$ from a selected lepton are removed. Jets are required to have $|\eta| < 2.1$ to ensure good $b$-tagging performance with the full jet contained within the Inner Detector acceptance. Jets containing $b$-quarks are identified using the decay length significance between the primary and a secondary vertex~\cite{SV0}. Jets with a significance greater than 5.85 are tagged as $b$-jets. In a simulated $t\bar{t}$ sample such a cut has a 50\% $b$-jet identification efficiency and a mis-tag rate of 10\% and 0.5\% for c and light jets respectively~\cite{SV0}. The $b$-tagging efficiency was measured in data using a $b$-jet enriched sample obtained by selecting jets containing muons, which are produced predominantly from $b$-hadron decay. The distribution of the muon transverse momentum relative to the jet axis, $p_T^{rel}$, is then used to discriminate between $b$, $c$- and light jets, before and after $b$-tagging.

\begin{table}[ht] 
\caption{Definition of the phase space for the 
 fiducial cross section for the W+$b$-jet and Z+$b$-jet measurements~\cite{refZb,refWb}. \label{tab:1} }
\begin{tabular}{lcc}
\hline\noalign{\smallskip}
Requirement & W + b & Z + b \\ 
\noalign{\smallskip}\hline\noalign{\smallskip}
Lepton $p_T$ & \multicolumn{2}{c}{$p_T^\ell > 20$~GeV} \\
Lepton $\eta$ & \multicolumn{2}{c}{$|\eta^\ell|<2.5$} \\
Dilepton mass & - & $76-106$~GeV \\
Neutrino $p_T$ & $p_T^\nu >25$~GeV & - \\
$W$ transverse mass & $m_T>40$~GeV & - \\
\hline
Jet $p_T$ & \multicolumn{2}{c}{$p_T^j>25$~GeV} \\
Jet rapidity & \multicolumn{2}{c}{$|y^j|<2.1$} \\
Jet multiplicity & $n \le 2$ & - \\
$b$-jet multiplicity & $n_b = 1$ or $n_b = 2$ & $n_b \geq 1$ \\
\hline
Jet-lepton separation & $\Delta R(\ell\mathrm{,jet}) > 0.5$ & - \\
\hline\noalign{\smallskip}
\end{tabular} 
\end{table}

\begin{figure}
\centering
\resizebox{0.8\columnwidth}{!}{%
\includegraphics{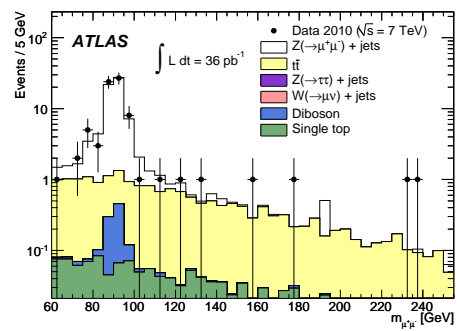} }
\caption{Di-lepton mass distribution for events with at least one b-tagged jet with $p_{T} > 25$GeV and $|y| < 2.1$ for the muon channel in the Z+$b$-jet analysis~\cite{refZb}. The contribution estimated from the simulated MC samples of the signal and various backgrounds is shown. The multi-jet background, estimated with a data-driven method, is not shown.}
\label{fig:A} 
\end{figure}

The estimated number of signal and background events in 35 pb$^{-1}$ of data is shown in Table.~\ref{tab:2}. The dilepton invariant mass in the muon channel for the Z+$b$-jet analysis is shown in Fig.~\ref{fig:A}. The dominant background is Z+jet events with a light or $c$-jet has been mistagged as a $b$-jet. Other backgrounds, including top and single-top, are small and therefore estimated from simulation. The small background from multi-jet production, referred to as QCD, was estimated from data, by fitting an exponential distribution in the dilepton mass in a QCD enriched sample obtained with a relaxed lepton selection criteria.

\begin{table}
\centering
\caption{Number of signal and background events in 35 pb$^{-1}$ of data~\cite{refZb,refWb}. The W+b numbers show the ALPGEN predictions normalised to the inclusive NNLO W cross-section, while the Z+jet yield has been normalised to the fit result. The QCD background in both cases has been estimated from data. The top background in the W+$b$-jet analysis was estimated using a partially data-driven method, while for Z+b it was estimated from simulation. The smaller single top and dibosons backgrounds were estimated from simulation. Uncertainties are not indicated.}
\label{tab:2} 
\begin{tabular}{lccc}
\hline\noalign{\smallskip}
 & Z + $\geq 1$~$b$-jet & W + 1 $b$-jet & W+ 2-jet \\
\noalign{\smallskip}\hline\noalign{\smallskip}
V + b & 64 & 43 & 45 \\
V + c & 60 & 192 & 81 \\
V + l & 0.0 & 68 & 37 \\
Top & 6 & 19.1 & 77 \\
Single Top & 0.3 & 31 & 41 \\
QCD & 1.0 & 18 & 15.8 \\
Diboson & 0.5 & 5.8 & 4.6 \\
\noalign{\smallskip}\hline
\end{tabular}
\end{table}

In the W+$b$-jet measurement, particularly in the 2-jet selection, the contribution from non W+jet backgrounds is large with a $S/B$ 0.1-0.2. Therefore both the QCD multijet and the top background were estimated from data. The top background was estimated in a control region with more than 4 jets and extrapolated into the signal region using Monte Carlo (MC) simulation. The uncertainty on the $b$-tagging efficiency cancels in this ratio therefore obtaining an estimate largely independent of the $b$-tagging uncertainty. The veto on additional jets was used to reduce the top background. 

\begin{figure}
\centering
\resizebox{0.8\columnwidth}{!}{%
\includegraphics{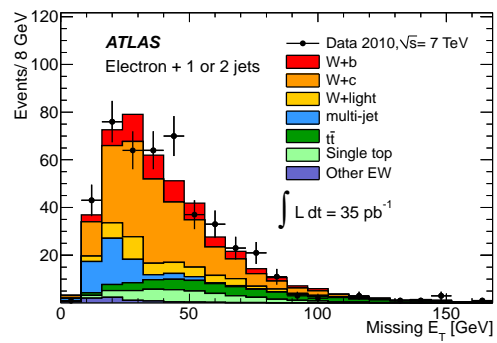} }
\caption{Missing ET distribution in the W+$b$-jet electron channel in the combined 1- and 2- jet bin after applying the $b$-tagging requirement~\cite{refWb}. All backgrounds, but QCD multi-jet, are normalised to their MC prediction.}
\label{fig:B} 
\end{figure}

Heavy-flavour jet production is the dominant process contributing to the QCD multi-jet background. As the QCD background is large and has a large uncertainty a very tight requirement on lepton isolation was introduced. In the muon channel the QCD background largely results from non-prompt muons and was extracted by exploiting the difference in efficiency between prompt and non-prompt muons to pass the standard selection criteria. In the electron channel the QCD background is estimated from a fit to the missing energy distribution with a template obtained by reversing certain electron identification criteria (see Fig.~\ref{fig:B}). The 50\% uncertainty on the QCD background estimate translates into a 7\% uncertainty on the W+$b$-jet cross-section. 

\begin{figure}
\centering
\resizebox{0.49\columnwidth}{!}{%
\includegraphics{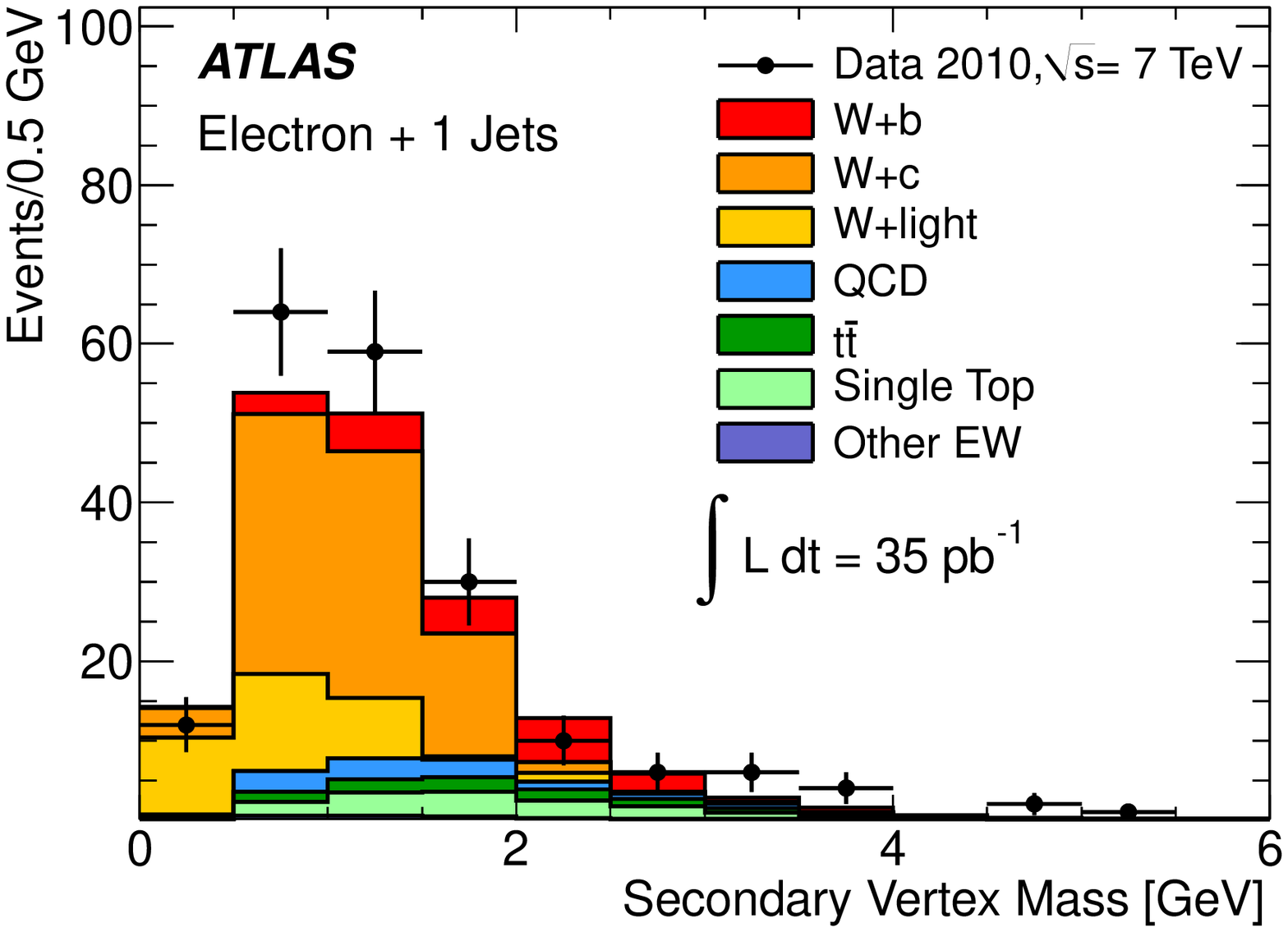} }
\resizebox{0.49\columnwidth}{!}{%
\includegraphics{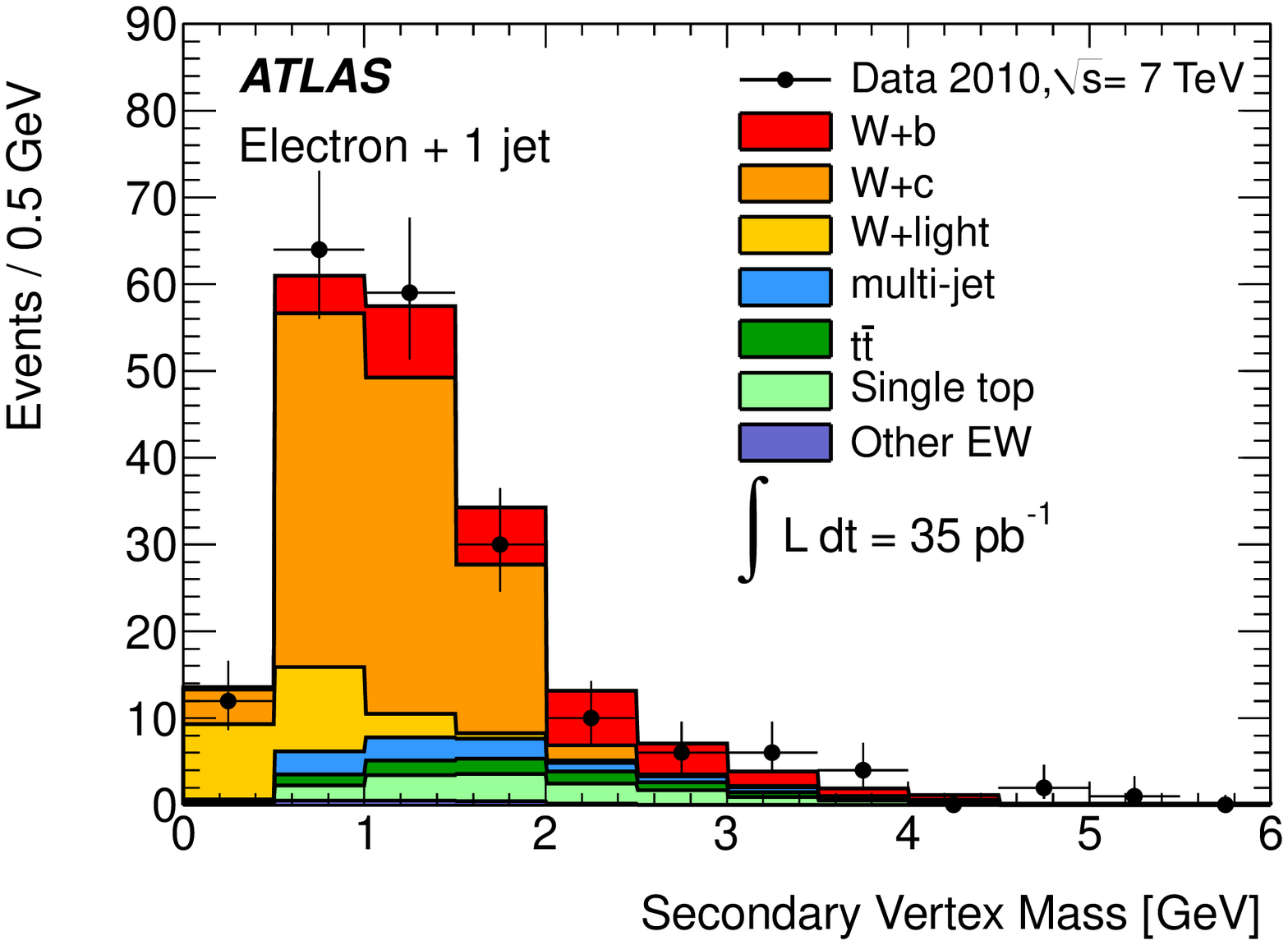} }
\caption{The secondary vertex mass distribution for b-tagged jets in W+$b$-jet events containing an electron and a single jet~\cite{refWb}. The distribution is shown with the $b$, $c$ and light contributions normalised to the inclusive NNLO W+jet cross-section (left) and to the fit results (right).}
\label{fig:1} 
\end{figure}

\begin{figure}
\centering
\resizebox{0.75\columnwidth}{!}{%
\includegraphics{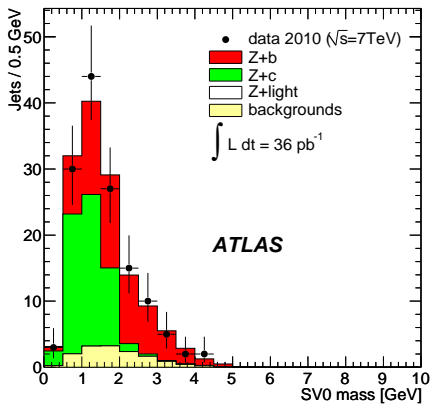} }
\caption{The secondary vertex mass distribution for b-tagged jets in selected Z+b events. The b, c and light templates have been normalised to the fit results. The small contribution from other backgrounds is also indicated~\cite{refZb}. }
\label{fig:2} 
\end{figure}

A maximum likelihood fit to the secondary vertex mass distribution was used to discriminate between $b-$, $c-$ and light-jets and to extract the flavour fraction on a statistical basis. The other non-V+jets backgrounds were allowed to float within uncertainties. The template shapes for each jet flavour were estimated from simulation and the systematic uncertainties on the shape were derived by comparing data and MC in multi-jet samples enriched in either light or $b$- and $c$-jets. Examples of the fits are shown in Fig.~\ref{fig:1} and~\ref{fig:2}. For the Z+$b$-jet measurement, the fit is performed for all $b$-jets in the electron and muon channels combined, while for the W+$b$-jet measurement it is performed separately by lepton flavour and jet multiplicity.

Figure~\ref{fig:1} shows the secondary vertex mass distribution for W+1-jet events. The distribution is shown both with the W+jet contribution normalised to the NNLO W cross-section (left) and after the W+jet contribution has been scaled by the fit results (right). The secondary vertex distribution for the Z+$b$-jet analysis is shown with the flavour contributions normalised to the fit results. There is no visible contribution from Z+light jets because the best fit value was consistent with zero within uncertainties.

The fiducial V+jet cross-sections were obtained by unfolding the measured $b$-fraction using MC simulation samples. Alpgen was used for the W+$b$-jet measurement and Sherpa for the Z+$b$-jet measurement. The measured cross-sections have uncertainties from the backgrounds, template shape and the unfolding factors.

Uncertainties on the $b$-tagging efficiency result in a systematic uncertainty of 10\% (12\%) for the Z(W)+$b$-jet measurement. The unfolding factor is sensitive to the accuracy of the Monte Carlo modelling such as the $b$-jet $p_T$ spectrum and uncertainty on the opening angle between the pairs of $b$-quarks. The $b$-jet $p_T$ spectrum is important because the $b$-tagging efficiency depends on $p_T$ and the opening angle distribution determines the probability that the pair of $b$-quarks will be reconstructed as a single jet. Uncertainties on the jet energy scale results in a 4\% (7\%) uncertainty in the Z(W)+$b$-jet measurement.

The results for the Z+$b$-jet measurement are summaris\-ed in Table~\ref{tab:3}. The Z+$b$-jet cross-section agrees well with the NLO MCFM prediction of 3.88 $\pm$ 0.58~pb~\cite{MCFM,MCFMZb}.

\begin{figure}
\centering
\resizebox{0.8\columnwidth}{!}{%
\includegraphics{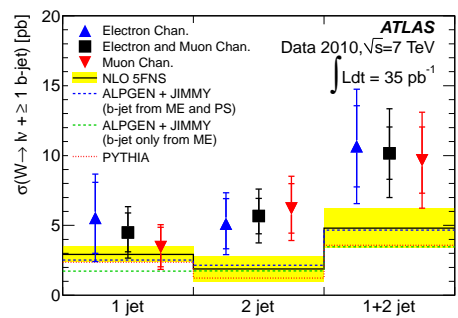} }
\caption{Comparison of the measured and predicted fiducial cross-sections of $b$-jet production in association with a W boson~\cite{refWb}. The cross-section is shown in exclusive bins for 1, 2 and 1+2 jets and separately for lepton flavours. The measurement is compared to NLO predictions obtained using the 5 flavour number scheme and to those from ALPGEN and Pythia.}
\label{fig:3} 
\end{figure}

The results for the W+$b$-jet measurement are summaris\-ed in Fig.~\ref{fig:3}. The measurements are compared to NLO predictions obtained using the 5 flavour number scheme~\cite{CampWb} after including a non-perturbative correction to the cross-section of $0.93 \pm 0.07$~\cite{refWb}. The correction includes a contribution from a $b\bar{b}$ pair in the final state with massive $b$-quarks and contribution with a massless $b$ quark in the initial state treated using a scheme based on $b$ quarks PDFs. The W+$b$-jet cross-section is systematically found to have a small excess over theoretical predictions, both in the 1 and 2-jet exclusive bins and the 1+2-jet exclusive bin. However the excess is small and the measurement is consistent at the 1.5~$\sigma$ level. 

\begin{table}
\centering
\caption{Measured and predicted fiducial cross-section of inclusive $b$-jet production in association with a Z-boson~\cite{refZb}}
\label{tab:3} 
\begin{tabular}{ll}
\hline\noalign{\smallskip}
 & Cross-section [pb] \\
\noalign{\smallskip}\hline\noalign{\smallskip}
Sherpa & $3.29 \pm 0.04$~(stat only) \\
ALPGEN & $2.23 \pm 0.01$~(stat only) \\
MCFM & $3.88 \pm 0.58$ \\
\hline\noalign{\smallskip}
ATLAS & $3.55^{+0.82}_{-0.74}$~(stat.) $\pm 0.12$~(syst) \\
\noalign{\smallskip}\hline
\end{tabular}
\end{table}

In conclusion, ATLAS measurements of vector boson in association with $b$-jets using the first 35~pb$^{-1}$ of data have been presented. Despite large uncertainties, the Z+$b$-jet cross-section is found to be consistent with NLO predictions, while a small excess at the 1.5$\sigma$ level is observed in the W+$b$-jet measurement.


\begin{thebibliography}{}
\bibitem{Hbb}
ATLAS Collaboration, ATLAS-CONF-2011-103, https://cdsweb.cern.ch/record/1369826/
\bibitem{SBottom}
ATLAS Collaboration, submitted to PRL, arXiv:1112.3832 
\bibitem{CampWb}
Campbell, J.M. et al, Phys. Rev. D. 79 (2009) 034023.
\bibitem{FFSWb}
Caola, F. et al, arXiv:1107.3714, (2011)
\bibitem{refZb}
ATLAS Collaboration, Phys. Lett. B 706 (2012) 295-313, arXiv:1109.1403
\bibitem{refWb}
ATLAS Collaboration, Phys. Lett. B 707 (2012) 418-437, arXiv:1109:1470
\bibitem{DetPaper}
ATLAS Collaboration, JINST 01 (2008), S08003
\bibitem{SV0}
ATLAS Collaboration, ATLAS-CONF-2011-089 (2011), https://cdsweb.cern.ch/record/1356198
\bibitem{MCFM}
Campbell, J.M. et al, Nucl.Phys.Proc.Suppl. 205-206 (2010) 10-15, arXiv:1007.3492 
\bibitem{MCFMZb}
Campbell, John M. et al. Phys.Rev. D69 (2004) 074021, hep-ph/0312024 






\end{thebibliography}
\end{document}